\documentclass[useAMS]{mn2e}

\begin{document}
\title[Theorem on gravitational wave backgrounds]
{A practical theorem on gravitational wave backgrounds}
\author[E.S. Phinney]{E.S.~Phinney\thanks{E-mail: esp@tapir.caltech.edu}\\
Theoretical Astrophysics,
130-33 Caltech,
Pasadena, CA 91125, USA}
\date{2001 July 31}
\pagerange{\pageref{firstpage}--\pageref{lastpage}}
\onecolumn    
\pubyear{2001} \volume{000}

\maketitle 
\label{firstpage} 
\begin{abstract}
There is an extremely simple relationship between the spectrum of
the gravitational wave background produced by a cosmological distribution
of discrete gravitational wave
sources, the total time-integrated
energy spectrum of an individual source,
and the present-day comoving number density of remnants.
Stated in this way, the background is entirely independent of the
cosmology, and only weakly dependent on the evolutionary history of
the sources. This relationship allows one easily to compute the
amplitude and spectrum of cosmic gravitational wave backgrounds from
a broad range of astrophysical sources, and to evaluate the uncertainties
therein.

\end{abstract}

\section{Introduction}
The strongest discrete sources of gravitational waves are those which
radiate large amounts of energy in a time (short compared to the age
of the universe) before or after a catastrophic event. Examples include
supernovae, accretion-induced collapse, black-hole formation events and
merging compact binary systems, which may involve white dwarfs, neutron
stars or black holes of the stellar or super-massive sort.  It is becoming
increasingly important to understand the `brightness' of the night sky
in gravitational waves due to the cosmic superposition of such sources.
This is crucial to the design of LISA, the NASA/ESA Laser Interferometer Space
Antenna, and for proposed follow-on missions at higher and lower frequencies
which might search for primordial stochastic backgrounds, e.g. from 
inflation or phase transitions in the early universe.
There have been a number of recent efforts to compute the backgrounds
from particular
classes of sources, by numerically integrating spectra or
waveforms of sources with assumed evolutionary histories 
over cosmological volumes \cite{coward,schneider,ferrari,rajagopal}.

We show
here that there is an extremely simple relationship between the spectrum of
the gravitational wave background produced by a cosmological distribution
of such sources and the {\em present-day comoving number density of remnants}.
The background is entirely independent of the cosmology, and in most cases
is almost independent of the evolutionary history of the sources.
This simple relation allows one to evaluate quickly uncertainties in
estimates of gravitational wave backgrounds and to survey new
backgrounds.  A companion paper \cite{phinney} in this way surveys the
gravitational wave sky, and points out a number of previously
unrecognized backgrounds of potential significance for
LISA and future missions.

In section \ref{sec:theorem} we give a simple physical derivation of 
the theorem. Section \ref{sec:example} applies the theorem to the
backgrounds produced by non-relativistic binaries in circular orbits,
and compares to previous calculations of the cosmic background from
double-degenerate binaries and merging super-massive black holes in
galactic nuclei. The reader wanting quick numbers for other circular orbit
sources should use equations \ref{eq:circOnumer}--\ref{eq:circSnumer}.
Section \ref{sec:mathematical} presents a
formal derivation of the theorem, and shows explicitly that it is valid
for all source lifetimes, including ones much longer and much
shorter than the duration of a measurement experiment.

\section{The Theorem: physical derivation}
\label{sec:theorem}

Let $f_r$ be the frequency of gravitational waves in a source's
cosmic rest frame, and $f$ the frequency of those waves observed today
on earth, $f_r=f(1+z)$. Let the 
total outgoing energy emitted in gravitational
waves between frequency $f_r$ and $f_r+df_r$ be
\begin{equation}
 \frac{dE_{gw}}{df_r}df_r
 \;.
\label{eq:dEdf}
\end{equation}
This energy, like $f_r$ is measured in the source's cosmic rest frame, and is
integrated over all solid angles and over the entire radiating lifetime
of the source. 

Let the number of events in unit comoving volume which occur
between redshift $z$ and $z+dz$ be $N(z)dz$. Define $\Omega_{gw}(f)$
to be the present-day energy density per logarithmic frequency interval, 
in gravitational waves of
frequency $f$, divided by the rest-mass energy density $\rho_c c^2$ that
would be required to close the universe.
Then the total present day energy density in gravitational
radiation is
\begin{equation}
{\mathcal E}_{gw}\equiv \int_0^\infty \rho_c c^2 \Omega_{gw}(f)\, {df/f}
 \equiv \int_0^\infty \frac{\pi}{4}\frac{c^2}{G} f^2 h_c^2(f)\frac{df}{f}
 \;,
\label{eq:edensdef}
\end{equation}
where $\rho_c=3H_0^2/(8\pi G)$ is the critical density of the universe, and
$h_c$ is the characteristic amplitude of the gravitational wave spectrum 
over a logarithmic frequency interval $d\ln f=df/f$. $h_c$ is related
to the one-sided ($0<f<\infty$) spectral density $S_{h,1}$ of the gravitational
wave background (cf. Thorne \shortcite{thorne}) by $h_c^2(f)=fS_{h,1}(f)$.
If the spectral density $S_{h,2}$ is defined to be
two-sided ($-\infty<f<\infty$,
cf. Ungarelli \&\ Vecchio \shortcite{ungarelli}, 
Cornish \shortcite{cornish}), the relation is $h_c^2(f)=2fS_{h,2}(f)$.
$S_h^{1/2}$ is often called the strain noise, since the mean square
signal strain output of an interferometer is simply the integral over all
frequencies of $S_h$ times the response function ${\mathcal R}(f)$
of the interferometer (of order unity for wavelengths much longer
than the interferometer --cf. section 3.2
of Cornish \& Larson \shortcite{cornish}).

In any homogeneous and isotropic universe,
the present-day energy density ${\mathcal E}_{gw}$ must be equal to sum of
the energy densities radiated at each redshift, divided by $(1+z)$ to account
for the redshifting of the gravitons since emission:
\begin{eqnarray}
{\mathcal E}_{gw} & \equiv & \int_0^\infty\int_0^\infty 
 N(z)\frac{1}{1+z}\frac{dE_{gw}}{df_r} f_r 
 \frac{df_r}{f_r}\, dz 
 \label{eq:edensourcea} \\
 & = & \int_0^\infty\int_0^\infty N(z)\frac{1}{1+z} 
        f_r\frac{dE_{gw}}{df_r} \, dz \, \frac{df}{f}
 \;. 
\label{eq:edensourceb}
\end{eqnarray}
Equating the two expressions in equations (\ref{eq:edensdef}) and
(\ref{eq:edensourceb}) for ${\mathcal E}_{gw}$ frequency by frequency, we find
\begin{equation}
 \rho_c c^2\Omega_{gw}(f)=\frac{\pi}{4}\frac{c^2}{G} f^2 h_c^2(f)
 =\int_0^\infty N(z)\frac{1}{1+z} 
 \left.\left(f_r\frac{dE_{gw}}{df_r}\right)\right|_{f_r=f(1+z)} 
 \, dz
 \;.
\label{eq:theorem}
\end{equation}
This theorem is our principal result. 
It has the simple physical interpretation that the energy density in 
gravitational waves per log frequency interval is equal to the comoving
number density of event remnants, times the (redshifted) energy each
event produced per log frequency interval.
Notice that the theorem does not depend
upon the cosmological model, except for the assumption of a homogeneous
and isotropic universe. Nor does it depend on any property of the
sources (beaming, polarization, etc) except for their 
time-integrated energy spectrum, provided they are randomly oriented with
respect to earth.
If more than one type of source is important, the right hand side
of equation \ref{eq:theorem} should be summed over the source types
$i$, i.e.
\begin{equation}
N(z)\left(f_r\frac{dE_{gw}}{df_r}\right) \mbox{ is replaced by } 
 \sum_i N_i(z)\left( f_r\frac{dE_{gw,i}}{df_r}\right)
 \;.
\end{equation}

The alert reader will detect some kinship between the frequency-integrated
version of this theorem (equation \ref{eq:edensourceb}) and
that of So\l tan \shortcite{soltan} relating the contribution of quasars to
the electromagnetic brightness of the night sky to the local space density
of remnant super-massive black holes. This relation becomes even closer
if the source we are considering is black holes or other compact
objects of mass $M_2$ gravitationally captured in circular orbits by 
super-massive black holes of mass $M_1\gg M_2$.
For them $\int (dE_{gw}/df_r)\, df_r=M_2c^2e_b$,
where $e_b=(e_{isco}-\int\dot e_H\,dt)$,
where $e_{isco}$ is the dimensionless binding energy per unit rest mass
of the innermost stable circular orbit (0.057 for Schwarzschild black holes,
0.42 and 0.038 respectively for prograde and retrograde equatorial orbits
about maximally rotating Kerr black holes), and $\int\dot e_H\,dt$ is the
(small) portion of the binding energy radiated down the black hole
horizon (note that this can be, and is, negative for prograde orbits 
around rapidly rotating black holes: the orbiting mass extracts some of
their rotational energy --see table VII of Finn \&\ Thorne 
\shortcite{finn}). The change
in the mass of the capturing black hole is 
$\Delta M_1=M_2(1-e_b)$. Thus 
\begin{equation}
{\mathcal E}_{gw}=\int_0^\infty N(z)\frac{1}{1+z}\frac{e_b}{1-e_b}\Delta M_1c^2\,dz
=\rho_\bullet c^2f_m\left\langle\frac{1}{1+z}\frac{e_b}{1-e_b}\right\rangle,
\end{equation}
where $f_m$ is the fraction of the present-day 
comoving mass density $\rho_{\bullet}$ in super-massive black holes which was
grown by the capture of compact objects (as opposed,
say to growth by axisymmetric
accretion of gas which results in negligible gravitational radiation). 

In section \ref{sec:mathematical} we give a more mathematical proof 
of the theorem, but we first give an example
of its use.

\section{An Example: Merging Binaries in Circular Orbits}
\label{sec:example}

In the Newtonian limit, a
circular binary of component masses $M_1$ and $M_2$ which merges due
to gravitational radiation losses in
less than the age of the universe has
\begin{equation}
\frac{dE_{gw}}{df_r}=\frac{\pi}{3}\frac{1}{G}
   \frac{(G{\mathcal M})^{5/3}}{(\pi f_r)^{1/3}} 
 \ \ 
   \mbox{for $f_{\min}<f_r<f_{\max}$}
 \;,
\label{eq:circE}
\end{equation}
where $\mathcal{M}$ is the chirp mass, 
$\mathcal{M}^{5/3}=M_1M_2(M_1+M_2)^{-1/3}$. Notice that to derive equation
\ref{eq:circE},
one need know nothing about gravitational radiation except that
it has twice the orbital frequency, and that the
orbital binding energy is removed by the gravitational radiation.

The lower limit $f_{\rm min}$ is set by the separation of the
system at its birth (or circularisation, whichever comes first), 
and our derivation applies only to those
systems whose initial
separation is small enough that their time to merge is much shorter
than the Hubble time at their birth.  
For example, a pair of $0.3M_{\sun}$ white dwarfs
must have $f_{\rm min}>10^{-4}\mbox{ Hz}$ in order to merge
through gravitational radiation in $<10^{10}\mbox{ y}$.
Over their lifetime, merging systems radiate a broad spectrum
of frequencies, up to an upper limit $f_{\rm max}$ set by the frequency 
at which the two bodies come into Roche lobe contact, or at which
tidal dissipation begins to dominate gravitational radiation in
determining the orbital evolution. For a binary
whose least massive star is a white dwarf of
mass $M_2=M_{wd} M_{\sun}$, one can use
the fit to white dwarf mass-radii of equation (17) of
Tout et al \shortcite{tout} and the approximation (good to better than
10\%) that the orbital frequency $\Omega_b=\pi f_r$ of a binary is
related to the equivalent volume radius of $M_2$'s Roche lobe $R_{L2}$
by $\Omega_b^2=0.1 GM_2/R_{L2}^3$, to find the maximum gravitational
wave frequency emitted before the onset of mass transfer:
\begin{equation}
 f_{\rm max}= 0.043 \frac{M_{wd}}{[1-(M_{wd}/1.44)^{4/3}]^{3/4}}\,\mbox{Hz}
 \;.
\end{equation}
For pairs of neutron stars, $f_{\rm max}\simeq 1.4\times 10^3\,\mbox{Hz}$
\cite{uryu,shibata}.
For small bodies spiraling into a non-rotating black hole of mass $M$,
$f_{\max}\simeq c^3/(6^{3/2}\pi GM) \simeq 4.3\times 10^3 (M/M_{\sun})^{-1}$
(for more detail and the generalization of equation \ref{eq:circE} to fully
relativistic orbits around rotating black holes, see 
Phinney \shortcite{phinney}).
Systems whose initial separation is so large
that their time to merge is much longer than the Hubble time at their birth
will have $f_{\rm max}\simeq f_{\min}$, and are neglected here.

Inserting equation \ref{eq:circE} into equation \ref{eq:theorem} gives
the gravitational wave background at $f<f_{\max}$
from a population of inspiraling binaries:
\begin{eqnarray}
\Omega_{gw}(f) & = & \frac{8\pi^{5/3}}{9}\frac{1}{c^2H_0^2}(G\mathcal{M})^{5/3}
  f^{2/3} N_0\langle(1+z)^{-1/3}\rangle \mbox{ and} 
  \label{eq:circtheoremO} \\
 h_c^2(f) & = & \frac{4}{3\pi^{1/3}}
  \frac{1}{c^2}\frac{(G\mathcal{M})^{5/3}}{f^{4/3}}
  N_0\langle(1+z)^{-1/3}\rangle
 \;,
\label{eq:circtheoremh}
\end{eqnarray}
where $N_0=\int_0^\infty N(z)\,dz$ is the present-day comoving
number density of merged remnants, and
\begin{equation}
\langle(1+z)^{-1/3}\rangle=\frac{1}{N_0}
  \int_{z_{\min}}^{z_{\max}}\frac{N(z)}{(1+z)^{1/3}}\,dz
 \;.
\label{eq:avgz}
\end{equation}
$z_{\min}=\max[0,f_{\min}/f-1]$ and $z_{\max}=f_{\max}/f-1$ 
can be set respectively to $0$ and $\infty$
except for $f$ just below $f_{\min}$ or $f_{\max}$.
In terms of the merger rate per comoving
volume $\dot N(t_r)$, $N(z)=\dot N dt_r/dz$, where
\begin{eqnarray}
\frac{dt_r}{dz} & = & \frac{1}{H_0(1+z)E(z)} \;,\mbox{ and}
 \label{eq:dtdz} \\
 E(z) & = & (\Omega_M(1+z)^3+\Omega_k(1+z)^2+\Omega_\Lambda)^{1/2} \;.
\label{eq:definee}
\end{eqnarray}
Currently favoured values are $\Omega_M=0.33$, $\Omega_\Lambda=0.67$,
$\Omega_k=0$, $H_0=65\mbox{ km s}^{-1}\mbox{Mpc}^{-1}$ 
(cf.\ Netterfield et al \shortcite{netterfield}). The value of
$\langle(1+z)^{-1/3}\rangle$ is not very sensitive to the details of
$N(z)$. For example, if we take $\dot N$ increasing rapidly
in the past, proportional to the
cosmic star formation rate as a function of redshift given in equation
(6) of Madau, Haardt \& Pozzetti \shortcite{madau} in a flat $\Omega_M=1$
universe, we have $\langle(1+z)^{-1/3}\rangle=0.74$, while a
time-independent $\dot N$ in a flat $\Omega_\Lambda=0.67$ universe
gives $\langle(1+z)^{-1/3}\rangle=0.80$.

The one approximation made in equations \ref{eq:circtheoremO} or
\ref{eq:circtheoremh} is that the lifetime has been assumed short
compared to the expansion time of the universe --i.e. all of the
inspiral occurs at the same redshift as the merger.  We are thus
neglecting the differential redshift of the very earliest parts
of the inspiral which occur at higher redshift than the merger.  But
because of the $t_{mrg}\propto a^4$ dependence on initial separation
$a$, sources generally either merge quickly or not at all, so this
approximation is poor only for a small percentage of sources with
finely tuned parameters.

In convenient numerical forms, equations \ref{eq:circtheoremO} and
\ref{eq:circtheoremh}
and the 1-sided strain noise $S_{h,1}^{1/2}(f>0)=h_c/\sqrt{f}$ 
(the two-sided 
$S_{h,2}^{1/2}(-\infty<f<\infty)$ is $S_{h,1}^{1/2}/\sqrt{2}$) become
\begin{eqnarray}
 \Omega_{gw}(f) &=& 1.3\times 10^{-17}
 \left(\frac{{\mathcal M}}{M_{\sun}}\right)^{5/3}
 \left(\frac{f}{10^{-3}\mbox{Hz}}\right)^{2/3}
 \left(\frac{N_0}{\mbox{Mpc}^{-3}}\right)
 \frac{\langle(1+z)^{-1/3}\rangle}{0.74}  \;,
 \label{eq:circOnumer} \\
h_c(f) &=& 3.0\times 10^{-24}
 \left(\frac{{\mathcal M}}{M_{\sun}}\right)^{5/6}
 \left(\frac{f}{10^{-3}\mbox{Hz}}\right)^{-2/3}
 \left(\frac{N_0}{\mbox{Mpc}^{-3}}\right)^{1/2}
 \left(\frac{\langle(1+z)^{-1/3}\rangle}{0.74}\right)^{1/2}  \;,
 \label{eq:circhnumer} \\
S_{h,1}^{1/2}(f) &=& 1.0\times 10^{-22}\mbox{Hz}^{-1/2}
 \left(\frac{{\mathcal M}}{M_{\sun}}\right)^{5/6}
 \left(\frac{f}{10^{-3}\mbox{Hz}}\right)^{-7/6}
 \left(\frac{N_0}{\mbox{Mpc}^{-3}}\right)^{1/2}
 \left(\frac{\langle(1+z)^{-1/3}\rangle}{0.74}\right)^{1/2}  \;.
 \label{eq:circSnumer}
\end{eqnarray}
The redshift averages in equations \ref{eq:circOnumer}--\ref{eq:circSnumer}
are frequency-independent constants for 
$f_{\min}/(1+z_*) \la f \la f_{\max}/(1+z_*)$,
where $z_*$ is the median source redshift; outside that range the 
edge effects of equation \ref{eq:avgz} become important.

We now use these to estimate the cosmic gravitational wave background
from merging double-degenerate binaries.  For our 
$H_0=65\mbox{ km s}^{-1}\mbox{Mpc}^{-1}$, Fukugita, Hogan \& Peebles
\shortcite{fukugita} give the mass density of stars in the universe
as $\rho_\ast=4.4\times 10^8M_{\sun}\mbox{Mpc}^{-3}$. After
$10^{10}\mbox{y}$ a stellar population with
the initial mass function of Kroupa \shortcite{kroupa} has 0.28 white
dwarf remnants per solar mass of stellar material, giving a present-day
comoving number density of $1.2\times 10^8 \mbox{WD Mpc}^{-3}$.
Of these white dwarfs, under standard assumptions of population
synthesis models (equal numbers of binaries per log initial
orbital separation, binary fraction 0.5, flat binary mass ratio
distribution), a fraction 0.015 of the white dwarfs will undergo
common envelope evolution leading to a pair of $\sim 0.3M_{\sun}$
white dwarfs which will merge in $<10^{10}\mbox{y}$. Thus we
estimate $N_0=1.8\times 10^6\mbox{Mpc}^{-3}$. Inserting this and
${\mathcal M}=0.26M_{\sun}$ into equations \ref{eq:circOnumer}
-- \ref{eq:circSnumer} and defining $f_{-3}=f/10^{-3}\mbox{Hz}$ gives
\begin{eqnarray}
 \Omega_{gw}(\mbox{wd-wd}) &=& 3\times 10^{-13} f_{-3}^{2/3} 
  \;, \nonumber \\
 h_c(\mbox{wd-wd}) &=& 1.4\times 10^{-21} f_{-3}^{-2/3} 
  \;,\nonumber \\
 S_{h,1}^{1/2}(\mbox{wd-wd}) &=& 4\times 10^{-20} f_{-3}^{-7/6}\mbox{Hz}^{-1/2}
 \;.
\label{eq:wdwdnumer}
\end{eqnarray}
These are respectively factors of 4, 2 and 2 less than found by the very
detailed calculations of Schneider et al \shortcite{schneider}. 
The difference is due largely
to the different choice of normalization and is representative of the
true uncertainties (see Phinney \shortcite{phinney} for an extensive
discussion): Schneider et al used the
Scalo \shortcite{scalo} IMF, a constant Galactic
supernova rate of $0.01\mbox{y}^{-1}$, and a binary fraction of 100 percent.
If we adopted the same binary fraction, and scaled their Galactic birthrate
($0.044\mbox{y}^{-1}$) to that of
Iben, Tutukov \& Yungelson \shortcite{iben} ($0.024\mbox{y}^{-1}$),
the numbers would be in perfect (but spurious) agreement.
The slight inaccuracies introduced by the theorem's
neglect of systems whose merger occurs
over a range of redshift (and our neglect of sources which exit
common envelope evolution at frequencies much higher than $10^{-4}\mbox{Hz}$)
are small compared to the astronomical uncertainties.

As a second example, we consider figure~4 of Rajagopal \& Romani
\shortcite{rajagopal}. They computed the gravitational wave background
due to mergers of super-massive black holes in galactic nuclei,
at the wave frequencies accessible through pulsar timing
$f\sim 0.1-1\mbox{y}^{-1}$.
Their assumptions are approximately that 0.2 of bright galaxies
contained a black hole, and that each bright galaxy had 5 mergers
in its lifetime, or on average one merger with another black hole. Their
black hole mass distribution has a black hole
space density $N_0\sim 10^{-4}\mbox{Mpc}^{-3}$ and
typical black hole mass of $\sim 10^{7.8}M_{\sun}$. Taking ${\mathcal M}$
for an equal mass pair, equation \ref{eq:circhnumer} gives
$h_c=8\times 10^{-17}(f/1 \mbox{y}^{-1})^{-2/3}$. However their mass function
has $N_0\propto M^{-0.4}$ up to $3\times 10^9M_{\sun}$, and
those heavier black holes merging with their typical $\sim 10^{7.8}M_{\sun}$
ones have ${\mathcal M}\propto M^{0.4}$, so the larger masses make
contributions to $h_c^2\propto M^{0.27}d\ln M$. Integrating over the mass
function gives a total
$h_c(f)\sim 2\times 10^{-16}(f/1 \mbox{y}^{-1})^{-2/3}$, in very good agreement
with the $f_{S1}=1$ curve in the simulation shown in their figure~4.
Of course, beliefs and data
about the numbers and mass distribution of super-massive
black holes have evolved considerably since 1995, so this value
is not the one preferred today (see Phinney \shortcite{phinney}).

It should be emphasized that in this discussion, we consider only
the typical amplitude and spectrum of the backgrounds. The important issue
of whether particular experiments could resolve
the various backgrounds (by identifying and
removing sources through optimal filtering of template waveforms, or
by angular resolution, or both) is deferred to later papers.

\section{The theorem again: mathematical derivation}
\label{sec:mathematical}
The flux of gravitational wave energy from a distant source is
\begin{equation}
 S(t)=\frac{c^3}{16\pi G}\left(\dot h_+^2+\dot h_{\times}^2\right)
 \;.
\label{eq:enfluxt}
\end{equation}
Denote the Fourier transforms of $h_+(t)$ and $h_{\times}(t)$ by
superscript tildes:
\begin{equation}
 \tilde h_{+,\times}(f)=\int_{-\infty}^{\infty}h_{+,\times}(t)e^{-i2\pi ft}\,dt
 \mbox{ with inverse } h_{+,\times}(t)=
 \int_{-\infty}^{\infty}h_{+,\times}(f)e^{i2\pi ft}\,dt
 \;.
\end{equation}
Apply Parseval's relation
\begin{equation}
 \int_{-\infty}^{\infty}\left|g(t)\right|^2\,dt=
 \int_{-\infty}^{\infty}\left|\tilde g(f)\right|^2\,df
\label{eq:parseval}
\end{equation}
to the time integral of equation \ref{eq:enfluxt}, using the fact that
the Fourier transforms of $\dot h_{+,\times}$ equal
$i2\pi f\tilde h_{+,\times}(f)$:
\begin{eqnarray}
 \int_{-\infty}^{\infty}S(t)\,dt & = &
 \frac{c^3}{16\pi G} \int_{-\infty}^{\infty}
   \left(\dot h_+^2+\dot h_{\times}^2\right)\,dt 
 \nonumber \\
 & = & \frac{\pi}{2}\frac{c^3}{G} \int_0^{\infty}
 f^2\left[ |\tilde h_+(f)|^2 + |\tilde h_{\times}(f)|^2\right]\,df
 \;,
\label{eq:enfluxone}
\end{eqnarray}
where in equation \ref{eq:enfluxone} we have made use of the fact that
$h_{+,\times}$ are real, 
$\tilde h_{+,\times}^\ast(f)=\tilde h_{+,\times}(-f)$, and hence
$|\tilde h_{+,\times}(f)|^2=|\tilde h_{+,\times}(-f)|^2$. This allows
us to fold the negative frequency part of the integral onto the
positive frequency part. If we average over source orientations
$\mathbf{\Omega}_s$, or equivalently, over observer positions around
a given source,
\begin{equation}
\langle S(t)\rangle_{\mathbf{\Omega}_s}=\frac{L_{gw}(t)}{4\pi d_L^2}
 \;,
\end{equation}
where $L_{gw}$ is the gravitational wave luminosity measured
in the cosmic rest frame
of the source, and $d_L$ is the luminosity distance to the source.
Since time $t$ at redshift $z=0$ is related to time $t_r$ in the
source's cosmic rest frame by $dt=(1+z)dt_r$ we have
\begin{equation}
 \int_{-\infty}^\infty \langle S(t)\rangle_{\mathbf{\Omega}_s} \,dt =
 \frac{1+z}{4\pi d_L^2} \int_{-\infty}^\infty L_{gw}(t_r)\, dt_r
 =\frac{1+z}{4\pi d_L^2} E_{gw}
 \;,
\label{eq:totalflux}
\end{equation}
where $E_{gw}$ is the rest-frame energy emitted in gravitational waves.

Comparing equations \ref{eq:enfluxone} and \ref{eq:totalflux}, and using
$df=(1+z)^{-1}df_r$ we get
\begin{eqnarray}
 \int_{-\infty}^\infty \langle S(t)\rangle_{\mathbf{\Omega}_s} \,dt & = &
 \frac{\pi}{2}\frac{c^3}{G}\frac{1}{1+z} \int_0^\infty
 f^2\left\langle |\tilde h_+(f)|^2+|\tilde h_{\times}(f)|^2
 \right\rangle_{\mathbf{\Omega}_s} \, df_r
 \nonumber \\
 & \equiv &
 \frac{1+z}{4\pi d_L^2}\int_0^\infty \frac{dE_{gw}}{df_r}\,df_r
 \label{eq:totaldedf}
 \;,
\end{eqnarray}
and thus identify
\begin{equation}
 \frac{dE_{gw}}{df_r}=\frac{2\pi^2 c^3}{G} d_M^2 f^2
 \left\langle |\tilde h_+(f)|^2+|\tilde h_{\times}(f)|^2 
 \right\rangle_{\mathbf{\Omega}_s}
 \;,
\end{equation}
where we have introduced the proper-motion distance $d_M=d_L/(1+z)$
(cf. section 5 of Hogg \shortcite{hogg}), 
which is also $1/(2\pi)$ times the proper (`comoving') circumference
of the sphere about the source which passes through the earth today.

Now consider sources undergoing the catastrophic events at redshift
$z$, at rate $\dot N$ per comoving volume per unit of cosmic time $t_r$
local to the event. As seen from earth, in earth time $dt$, the number
of events which occur in $dt$ between redshift $z$ and $z+dz$ is
\begin{equation}
 \frac{d \#}{dt\,dz}=\dot N\frac{1}{1+z}\,\frac{d{\mathcal V}_c}{dz}
 \;,
 \label{eq:numbevents}
\end{equation}
where the comoving volume element is (cf. Hogg \shortcite{hogg} and
references therein)
\begin{equation}
 \frac{d{\mathcal V}_c}{dz}=4\pi \frac{c}{H_0}d_M^2\frac{1}{E(z)}
 \;,
\label{eq:comovingvol}
\end{equation}
where $E(z)$ was defined in equation \ref{eq:definee}.
The number of events which occur in a comoving volume between
the cosmic times $t_r(z)$ and $t_r(z+dz)$ is
\begin{equation}
 N(z)=\dot N\frac{dt_r}{dz}=\dot N\frac{1}{(1+z)H_0 E(z)} \;.
\end{equation}
Thus equation \ref{eq:numbevents} can be rewritten
\begin{equation}
 \frac{d\#}{dt\,dz}=N(z) c4\pi d_M^2  \;.
\end{equation}
The energy density in gravitational waves at $z=0$ is then, inserting
equation \ref{eq:totaldedf}
\begin{eqnarray}
{\mathcal E}_{gw}=\int \Omega_{gw}(f)\rho_c c^2 df/f & = &
 \int_0^\infty\frac{1}{c}
 \left[ \int_{-\infty}^\infty \langle S(t)\rangle_{\mathbf{\Omega}_s}\,dt
 \right]
 \frac{d\#}{dt\,dz}\,dz  
 \label{eq:mathresult} \\
 & = & \int_0^\infty \frac{1+z}{4\pi d_L^2 c}
 \left[ \int_0^\infty \left. f_r \frac{dE_{gw}}{df_r}\right|_{f_r=f(1+z)}
 \frac{df}{f}\right] N(z) c 4\pi d_M^2 \,dz  \nonumber \\
 & = &
 \int_0^\infty \int_0^\infty N(z)\frac{1}{1+z}
 \left.\left(f_r\frac{dE_{gw}}{df_r}\right)\right|_{f_r=f(1+z)}
  \,dz\,\frac{df}{f}  \;,
 \label{eq:edenagain}
\end{eqnarray}
which reproduces equation \ref{eq:edensourceb} and hence 
the statement of the theorem, equation \ref{eq:theorem}.

Notice that equation \ref{eq:mathresult} and the result \ref{eq:edenagain}
are {\em independent} of whether the timescale
over which an individual source emits gravitational waves at the frequencies
of interest is long or short compared to the observing time.
The case when it is long compared to the observing time is
appropriate e.g. for the two examples treated in section \ref{sec:example}:
double degenerate binaries in the LISA frequency band, 
lifetimes $10^2$--$10^6\mbox{y}$, and
the super-massive binaries contributing to the pulsar timing background,
lifetimes $10^4$--$10^6\mbox{y}$.
In this case the number of sources contributing to the background is 
much larger than the
number of events (binary mergers in these cases) in the observing time $T$,
by the ratio $E_{gw}/(L_{gw}T)$. But the energy which each radiates in
$T$ is only a fraction $(L_{gw}T)/E_{gw}$ of the total given by equation
\ref{eq:totalflux}. So the product of the number of sources and the
energy (spectrum) radiated by each is independent of $T$ and $L_{gw}$.
This must be so, since ${\mathcal E}_{gw}$ is independent of
whether the observing time is $1\mbox{y}$ or $10^9\mbox{y}$.

The observational prospects for {\em removing} the background by fitting
individual sources {\it do} depend on the observing time, however, so
it is instructive to re-derive equations \ref{eq:circtheoremO} and
\ref{eq:circtheoremh} for the case of merging binaries in circular
orbits, keeping track of the number of sources contributing in each
frequency interval, while explicitly showing the cancellation discussed
in the previous paragraph. 

For circular binaries of separation $a$, masses $M_1$ and $M_2$ and
quadrupole gravitational wave frequency $f_r=2/P_b$ given by
\begin{equation}
 G(M_1+M_2)=\pi^2 f_r^2 a^3 \;,
 \label{eq:gwfreq}
\end{equation}
and the energy flux at earth averaged over source orientations and orbital
phases is
\begin{equation}
 S=\frac{\pi}{4}\frac{c^3}{G}\frac{2}{5}
 \left(\frac{4 G^2M_1M_2}{c^4 a d_M}\right)^2
 \left(\frac{f_r}{1+z}\right)^2 =\frac{L_{gw}}{(1+z)^2 d_M^2} \;.
 \label{eq:circflux}
\end{equation}
If the rate of mergers per comoving volume is $\dot N$, then the
space density $N(a)da$ of systems with separations between $a$ and
$a+da$ is given by the continuity equation
\begin{equation}
 \frac{d}{da}\left(\dot a N(a)\right)=-\dot N \delta(a) \;,
\end{equation}
where we have assumed that most of the sources are born at much
larger separations than those of interest. Then $N(a)=\dot N/(-\dot a)$.
Since
\begin{equation}
 \frac{d}{dt}\frac{GM_1M_2}{2a}=L_{gw} \;,
\end{equation}
one finds
\begin{equation}
 \dot a=-\frac{2 a^2 L_{gw}}{GM_1M_2} \mbox{ and } 
 N(a)=\dot N \frac{GM_1M_2}{2a^2L_{gw}} \;.
 \label{eq:adot}
\end{equation}
The number density of sources $N(f_r)df_r$ radiating at frequencies between
$f_r$ and $f_r+df_r$ is related to $N(a)$ by $-N(a)da=N(f_r)df_r$, which gives,
after manipulation
\begin{equation}
 N(f_r)=\frac{5\pi}{96}\frac{c^5}{(G{\mathcal M})^{5/3}}
 \frac{\dot N}{(\pi f_r)^{11/3}} \;.
\label{eq:noffr}
\end{equation}
The contribution to the specific intensity $I$ 
($\mbox{erg cm}^{-2}\mbox{s}^{-1}\mbox{sr}^{-1}\mbox{Hz}^{-1}$) of
the gravitational waves from sources between $z$ and $z+dz$ is
\begin{equation}
dI=S \cdot N(f_r) \frac{d{\mathcal V}_c}{d\Omega dz}\frac{df_r}{df}\,dz \;.
\end{equation}
Integrating over all redshifts gives the total specific intensity.
Into this equation, insert the right side of equation \ref{eq:circflux}
for $S$. For $N(f_r)$ use $N(a)da/df_r$, with $N(a)$ from equation
\ref{eq:adot}; for $d{\mathcal V}_c/dzd\Omega$ use 
$1/(4\pi)$ times equation \ref{eq:comovingvol}, and recall that
$df_r/df=1+z$. After using also equation \ref{eq:dtdz}, one gets
\begin{equation}
 I=\frac{1}{4\pi}\int_0^\infty c \left(\dot N \frac{dt}{dz}\right)
 \frac{GM_1M_2}{2a^2}
 \frac{-da}{df_r}\,dz\;.
\end{equation}
The quantity in parenthesis is $N(z)$.
The energy density per log (observed) frequency is, using
$f=f_r/(1+z)$ in the second line,
\begin{eqnarray}
 \Omega_{gw}(f)\rho_c c^2=\frac{f}{c}\int I\,d\Omega
 & = & \frac{4\pi}{c}f I \nonumber \\
 & = & \int_0^\infty \frac{N(z)}{1+z} \frac{GM_1M_2}{2a}
      \left(\frac{-d\ln a}{d\ln f_r}\right) \,dz \nonumber \\
 & = & \frac{2}{3}\int_0^\infty \frac{N(z)}{1+z} \frac{GM_1M_2}{2}
      \left[\frac{\pi^2 f_r^2}{G(M_1+M_2)}\right]^{1/3}
  \nonumber \\
 & = & \frac{\pi^{2/3}}{3}\frac{(G{\mathcal M})^{5/3}}{G}f^{2/3}
      \int_0^\infty \frac{N(z)\,dz}{(1+z)^{1/3}}\;.
\end{eqnarray}
The final form reproduces equation \ref{eq:circtheoremh}, while
equation \ref{eq:noffr} allows one to check the instantaneous
number of sources contributing to a given frequency interval of the
background.

\section{Conclusion}
We have shown that in order to estimate the contribution to
the gravitational wave background from a population of sources, one
need not specify a cosmology.
The only work involved is some relativity (often simple):
computing $f_r dE{gw}/df_r$ for the life-history of an individual source,
and some astronomy: estimating the number of sources which have ever
lived and died in a unit comoving volume of the universe, and some
idea of at what redshift they did so. Insert the results in equation
\ref{eq:theorem} to get the full spectrum of the background.

\section*{Acknowledgments}
This work was supported in part by NASA grants NAG5-7034 and
NAG5-10707.

\label{lastpage}
\bsp
\end{document}